\documentclass[12pt]{article}
\usepackage{amssymb,amsmath,amscd}
\usepackage{amsthm}

\newtheorem{theorem}{Theorem}[section]

\newtheorem{lemma}[theorem]{Lemma}

\date{}

\begin{document}
\author{ Evgeny Lakshtanov\thanks{Department of Mathematics, Aveiro University, Aveiro 3810, Portugal.  This work was supported by Portuguese funds through the CIDMA - Center for Research and Development in Mathematics and Applications and the Portuguese Foundation for Science and Technology (``FCT--Fund\c{c}\~{a}o para a Ci\^{e}ncia e a Tecnologia''), within project PEst-OE/MAT/UI4106/2014 (lakshtanov@ua.pt).} \and
 Boris Vainberg\thanks{Department
of Mathematics and Statistics, University of North Carolina,
Charlotte, NC 28223, USA. The work was partially supported  by the NSF grant DMS-1410547 (brvainbe@uncc.edu).}}

\title{Exceptional points in Faddeev scattering problem}


\maketitle

\begin{abstract}
 Exceptional points are values of the spectral parameter for which the homogeneous Faddeev scattering problem has a non-trivial solution.
We study the existence/absence of exceptional points for small perturbations of conductive potentials of arbitrary shape and show that problems with absorbing potentials do not have exceptional points in a neighborhood of the origin. A criterion for existence of exceptional points is given.
\end{abstract}


\section{Introduction}
The paper concerns the 2-D Faddeev scattering problem where incident waves grow exponentially at infinity \cite{faddeev}. This problem is used often for solving inverse problems \cite{Nachman}-\cite{RN2}. Let us recall the statement of the Faddeev problem. Let $\mathcal O$ be an open bounded domain in $\mathbb R^2=\mathbb R^2_z,~z=(x,y),$ with $C^2$ boundary $\partial \mathcal O$ and the outward normal $\nu$.

Let $\zeta=(\zeta_1,\zeta_2)\neq 0$ be a vector with complex components $\zeta_i\in\mathbb C$ and $\zeta_1^2+\zeta_2^2=E\geq 0$. Mostly we will deal with the case when the energy $E$ is equal to zero, i.e., $\zeta=(k,\pm ik),~k\in \mathbb C \backslash 0 $.  Then $u=u(z,\zeta)$ is a solution of the Faddeev scattering problem if
\begin{equation}\label{ref0401B}
-\Delta u - n u =0,   ~ z\in \mathbb R^2,
\end{equation}
\begin{equation}\label{ref0401Ba}
u(z,\zeta)=e^{i\zeta \cdot z} + u^{out}, \quad e^{-i\zeta \cdot z} u^{out} \in W^{1,p}(\mathbb R^2), ~ p>2,
\end{equation}
where $\zeta\cdot z=\zeta_1x+\zeta_2y$ and the potential $n=n(z)$ is bounded in $\mathcal O$, complex-valued if the opposite is not claimed, and vanishes outside $\overline{\mathcal O}$.

The main object of our study is the set $\mathcal E$ of {\it exceptional points} $ 0\neq k\in \mathbb C$ such that the homogeneous problem (\ref{ref0401B}), (\ref{ref0401Ba}) with $ \zeta=(k,\pm ik)$ has a nontrivial solution. This homogeneous problem has the form
\begin{equation}\label{ref0414A}
-\Delta v- n v =0,   ~   z\in \mathbb R^2; \quad
e^{-i\zeta \cdot z} v\in W^{1,p}(\mathbb R^2), ~ p>2.
\end{equation}
We will say that an exceptional point has {\it multiplicity} $m$ if the dimension of the solution space of problem (\ref{ref0414A}) is $m$.  We restrict ourself to the case $\zeta=(k,ik), k \in \mathbb C, $ the case $\zeta=(k,-ik), k \in \mathbb C \backslash \{0\}, $ can be treated similarly (see discussion in \cite[p.76]{Nachman}).

In the case of positive energy, equation (\ref{ref0401B}) should be replaced by
\begin{equation}\label{ref0408B}
-\Delta u -Eu- n u =0,   ~ z=(x,y)\in \mathbb R^2.
\end{equation}
 The following parametrization (see \cite{RN2}) of $\zeta\in \mathbb C^2,~\zeta^2=E>0,$ is used in this case instead of $\zeta=(k,ik)$:
$$
\zeta = \left ( \begin{array}{c}
(\lambda + \frac{1}{\lambda})\frac{\sqrt{E}}{2} \\
(\frac{1}{\lambda} - \lambda)i\frac{\sqrt{E}}{2}
\end{array} \right ), \quad |\lambda| \neq 1.
$$
The fundamental solution that corresponds to outgoing waves governed by (\ref{ref0408B}) has the form
$$
G_\zeta(z)= e^{i\zeta \cdot z}  \frac{1}{(2\pi)^2} \int_{\mathbb R^2} \frac{e^{iz' \cdot z}}{|z'|^2  + 2 \zeta \cdot z' } dz' ,
 $$
and condition (\ref{ref0401Ba}) should be replaced by the representation of the solution $u$ through $G$:
\begin{equation}\label{2a}
u(z,\zeta)=e^{i\zeta \cdot z} + \int_{\partial \mathcal O} G_\zeta(z-w)\mu_\zeta(w)dl_w, ~ \mu \in H^{-\frac{1}{2}}(\partial \mathcal O), ~ z \in \mathbb R^2 \backslash \mathcal O.
\end{equation}

{\bf Definition.} A point $\lambda \in \mathbb C \backslash \{0\}, |\lambda|\neq 1,$ will be called {\it exceptional} if problem (\ref{ref0408B}), (\ref{2a}) has a nontrivial solution. The multiplicity of an exceptional point is defined by the number of linearly independent solutions of (\ref{ref0408B}), (\ref{2a}). The set of all exceptional points will be denoted by $ \mathcal E(E)$.

Note that the incident waves in the classical scattering problem (with positive energy) have the form $e^{i(k_1x+k_2y)}$. The exceptional set in this case is empty due to the absence of eigenvalues imbedded into continuous spectrum (a very simple proof can be found in \cite{vai}). Similar arguments do not work for the Faddeev scattering problem since outgoing solutions of (\ref{ref0414A}) and (\ref{ref0408B}) do not decay at infinity.

The knowledge of exceptional points is particularly important when the Faddeev scattering problem is applied to solve the inverse problem of recovering the potential $n$ from the Dirichlet-to-Neuman map $F_n$ at the boundary $\partial\mathcal O$, which is defined by solutions of either equation (\ref{ref0401B}) or (\ref{ref0408B}) in $\mathcal O$. For example, if the real potential is continued by zero in the exterior of $\mathcal O$, then the following relation holds for the solution $u$ of (\ref{ref0401B}) under the condition of absence of exceptional points (e.g., see \cite{Nachman} and the discussion after formula (21) in \cite{music})
\begin{equation}\label{dbarinteq}
e^{-i\zeta \cdot z}u(z,k)=1- \frac{1}{(2\pi)^2} \int_{\mathbb R^2} \frac{t(k')}{(k'-k)\overline{k}'}e_{-z}(k') \overline{e^{-i\zeta' \cdot z}u(z,k')}dk'_1 dk'_2.
\end{equation}\label{inte}
Here $e_z(k)=\exp((kz+\overline{k}\overline{z}))$, and the
coefficient $t(k)$ is the so-called direct scattering transform that could be calculated through $F_n$:
\begin{equation}\label{dst}
t(k)=\int_{\partial \mathcal O} e^{i\overline{kz}}[(F_n-F_0)u](z,k) dl_z,
\end{equation}
Function $u$ on $\partial\mathcal O$ in (\ref{dst}) can be found without using the potential (see formula (\ref{LS}) below).

In order to complete the solution of the inverse problem, one also needs to know that the solution of the integral equation (\ref{dbarinteq}) is unique. Then the potential $n(z)$ can be found using the approach developed in  \cite{Nachman} or just as $-\frac{\Delta u}{u}$, see \cite{beals},\cite{Henkin}.

This inverse method was justified in the case of conductive potentials \cite{Nachman}. The latter potentials have the form $n=-q^{-\frac{1}{2}}\Delta q^{\frac{1}{2}}$, where $q$ is smooth, non-negative, and $q-1$ vanishes outside $\mathcal O$. In this case, equation (\ref{ref0401B}) can be reduced to the equation $\nabla (q \nabla v)=0$ by the substitution $u=\sqrt{q}v$. The case of so-called subcritical potentials was studied in \cite{music}.

Exceptional points for sign definite perturbations of conductive potentials were studied in \cite{siltanen2}, \cite{siltanen3} under the condition that the potential and its perturbation are spherically symmetric. The existence or non-existence of exceptional points in this case depends on the sign of the perturbation.

Similarly, one can extend the potential to the exterior of $\mathcal O$ by a nonzero real constant $E$. The absence of exceptional points and the uniqueness of solutions of the corresponding integral equation was justified in the case of small enough potentials (see \cite{grinevich}), and for any potential if $E$ is large enough (see \cite{RN2}).  Let us also note that an integral equation similar to (\ref{dbarinteq}) was obtained in certain cases \cite[(8.27),(8.28)]{RN2} under the condition that only a neighborhood of the origin and a neighborhood of infinity are free of the exceptional points.

It is worth noting that the location of exceptional points is time independent for potentials that satisfy the Novikov-Veselov equation (eg. \cite{bogdanov},\cite{siltanen4}). The latter is a multidimensional generalization of the KdV equation.

One can find the solution $u(z,k)$ of (\ref{ref0401B}), (\ref{ref0401Ba}) by reducing the problem to the Lipmann-Schwinger equation, which leads to (see e.g. \cite{RN})
\begin{equation}\label{LS}
u(z,k)=(I+S_k(F_n-F_0))^{-1}e^{i\zeta \cdot z}, ~ z \in \partial \mathcal O.
\end{equation}
Here $F_n$ is the Dirichlet-to-Neumann map for the equation $(-\Delta-n)u=0$ in $\mathcal O, ~~F_0=F_n|_{n=0}$, $S=S_k$ is the single layer operator on the boundary with Faddeev's Green function $G_k(z)$:
\begin{equation}\label{skk}
S_k:H^{-\frac{1}{2}}(\partial \mathcal O)\to H^{\frac{1}{2}}(\partial \mathcal O);~~ \quad S_k\sigma(z)=\int_{\partial \mathcal O}G_k(z-z')\sigma(z')dl_{z'},~~z\in \partial \mathcal O,
\end{equation}
where $dl$ is the element of the length and
\[
G_k(z)=\frac{1}{(2\pi)^2} e^{i\zeta \cdot z} \int_{\mathbb R^2} \frac{e^{i(\xi_1x+\xi_2 y)}}{|\xi|^2+2k\xi}d\xi_1 d\xi_2, \quad \xi=\xi_1+i\xi_2.
\]
Function $G_k$ is real-valued (see e.g. \cite[Part 3.1.1]{sthesis}). Indeed, the second condition in (\ref{ref0414A}) can be written in the form $|e^{i\zeta \cdot z}|u^{out} \in W^{1,p}(\mathbb R^2)$. From here it follows that $\Re G_k$ is the Green function, while $\Im G_k=0$ due to Nachman's uniqueness result \cite{Nachman}.

Equation (\ref{LS}) provides the standard basis for studying the exceptional set $\mathcal E$: exceptional points can be defined as the values of the parameter $k\in \mathbb C \backslash \{0\} $ for which the non-self-adjoint family of operators
\begin{equation}\label{expo}
I+S_k(F_n-F_0)
\end{equation}
has a non-trivial kernel. It is natural to consider {\bf $S_k(F_n-F_0)$} as operator in $L_2(\partial \mathcal O)$ or in Sobolev spaces $H^{s}(\partial \mathcal O)$, where value of $s$ is restricted only by smoothness of $\partial\mathcal O$. The second term in  (\ref{expo}) is a compact operator in each of these spaces, and the kernel of operator (\ref{expo}) does not depend on the choice of the space.

{\bf Description of the results obtained below}. Our progress in the study of $\mathcal E$ is based on establishing a connection between exceptional values of $k$ and the kernels of the family of operators $F_n-F^{out}(k)$, which are easier to control than the kernels of operator (\ref{expo}).
Here $F^{out}(k)$ is the Dirichlet-to-Neumann map for the exterior Faddeev scattering problem in $\mathbb R^2 \backslash \mathcal O$. The next section starts with studying some properties of operator $F^{out}(k)$. Then we prove that a point $k\in \mathbb C \backslash \{0\} $ is exceptional if and only if the kernel of the  family of operators $F_n-F^{out}(k)$ is nontrivial. Moreover, the dimension of the kernel coincides with the multiplicity of the exceptional point (which is the dimension of the solution space of problem (\ref{ref0414A})).

Section 3 contains the two main results of this paper. They follow from this new criterion for the exceptional points. The first result is the absence of exceptional points in a neighborhood of the origin $k=0$ in the case of absorbing potentials. The second one is a generalisation of the results  of  \cite{siltanen2}, \cite{siltanen3} on sign definite perturbations of conductive potentials to non spherically symmetrical problems. We do not assume the radial symmetry of either the underlying conductive potential or its perturbation. We prove the absence of exceptional points for perturbations of a specific sign and the existence of the exceptional set for perturbations of the opposite sign. This set is close to a circle centred at the origin. A criterion for existence of exceptional points is given in the last short section.

\section{Reduction to boundary operators. }

We consider the Faddeev scattering problem (\ref{ref0401B}), (\ref{ref0401Ba}) with zero energy in this section.
Without loss of the generality we can assume that the equation $-\Delta v- n v =0$ in $\mathcal O$ does not have non-trivial solutions vanishing at the boundary (i.e., zero is not an eigenvalue of the interior Dirichlet problem). In other words, the operator $F_n$ is well defined. We can make sure that this condition holds by extending the domain $\mathcal O$ slightly while preserving the function $n$.

Consider the exterior Dirichlet problem
\begin{equation}\label{ccc1}
-\Delta u =0,   ~ z\in \mathbb R^2\backslash \overline{\mathcal O}; \quad u|_{\partial \mathcal O}=f \in H^{ \frac{1}{2}}(\partial \mathcal O);
 \quad e^{-i\zeta \cdot z} u \in W^{1,p}(\mathbb R^2 \backslash \overline{\mathcal O}), ~ p>2.
\end{equation}
By $F^{out}(k):H^{\frac{1}{2}}(\partial \mathcal O)\to H^{-\frac{1}{2}}(\partial \mathcal O)$ we denote the operator that maps the Dirichlet data $f$ into the outward (with respect to $\mathcal O$) normal derivative $u_\nu$ of the solution of the problem (\ref{ccc1}) at the boundary $\partial \mathcal O$. Denote by $\mathcal E_D$ the set of values of $k\in \mathbb C \backslash \{0\}$ for which the homogeneous problem (\ref{ccc1}) has a non-trivial solution (the subindex $D$ here stands for the Dirichlet).

{\it Definition.} We will call a set $\{k=k_1+ik_2\}\subset\mathbb C$ {\it a 1-D real analytic variety} if the set of corresponding points $(k_1,k_2)\in \mathbb R^2$ is an intersection of a 1-D analytic variety in $\mathbb C^2=\mathbb C^2_{k_1,k_2}$ with the Euclidean space $\mathbb R^2$. Let us stress the meaning of the notation $k$. It will be used for points $k=k_1+ik_2$ of the complex plane. When we need to think about these points as vectors in the Euclidian space $\mathbb R^2$, we will use notation $(k_1,k_2)$ instead of $k$.

{\it Definition}.  We will say that an operator $A$ has a real-valued integral kernel if $\overline{Af}=A\overline{f}$ for every function $f$ from the domain of $A$.
The operator $A^\dag:=(A-A^*)/{2i}$ is called the non-self-adjoint part of $A$.

The following lemma concerns the exterior Faddeev problem (\ref{ccc1}).  Let us introduce the following parameter $\varepsilon=\varepsilon(k):=[-\nu(\frac{\gamma}{2\pi}+\frac{1}{2\pi}\ln|k|)]^{-1},~|k|\ll1,$ where $\gamma$ is the Euler constant
 and $\nu=|\partial\mathcal O|$ is the boundary length.
 \begin{lemma}\label{lemma0408C} 1) The set $\mathcal E_D\subset \mathbb C$ is a real analytic variety and coincides with the set $\mathcal K$ of values of $0\neq k\in \mathbb C$ for which the operator $S_k$ has a non-trivial kernel. The operator $S_k$ is onto when $k\notin \mathcal E_D\bigcup\{0\}$.

2) The Dirichlet-to-Neumann map
$$
F^{out}(k):H^{\frac{1}{2}}(\partial \mathcal O)\to H^{-\frac{1}{2}}(\partial \mathcal O), \quad  k\notin \mathcal E_D\bigcup\{0\}
$$
of the exterior Faddeev problem (\ref{ccc1}) exists, is analytic in each of the variables $k_1,k_2$, and is an elliptic pseudo-differential operator  of the first order with negative symbol.

 3) Operator
$F^{out}(k): H^{\frac{1}{2}}(\partial \mathcal O)\to H^{-\frac{1}{2}}(\partial \mathcal O), ~k\neq 0,$  admits a continuous extension at point $k=0$. When $|k|$ is small enough, the extended operator is an infinitely smooth function of $\varepsilon:=[-\nu(\frac{\gamma}{2\pi}+\frac{1}{2\pi}\ln|k|)]^{-1}\geq 0$ and $\arg k$, and it has the following properties:

Operator $F^{out}(0)$ is self-adjoint, has isolated simple eigenvalue $\lambda=0$ with constant eigenfunctions, and there exists $\delta>0$ such that $F^{out}(0)\leq -\delta<0$ on the subspace $ H^{\frac{1}{2},\bot}(\partial \mathcal O)$.

4) Operator $F^{out}(k), ~k\notin \mathcal E_D\bigcup\{0\},$  has a real-valued integral kernel. The non-self-adjoint part of $F^{out}(k), ~k\notin \mathcal E_D\bigcup\{0\},$  is a smoothing operator whose norm vanishes as $|k|\to 0$. To be more exact,
\begin{equation}\label{gl}
\|(F^{out})^\dag\varphi\|_{H^{1/2}(\partial \mathcal O)}\leq C|k|\|\varphi\|_{H^{-1/2}(\partial \mathcal O)}.
\end{equation}
\end{lemma}
{\bf Proof.}
We will start with a study of invertibility of operator $S_k$ (defined by (\ref{skk})) as $k\to 0$. In particular, we are going to prove that the set $\mathcal K$ (of values of $0\neq k\in \mathbb C$ for which the operator $S_k$ has a non-trivial kernel) is a real analytic variety. Later we will show that set $\mathcal K$ coincides with $\mathcal E_D$, and therefore the latter is also a real analytic variety. Our first step is to show that $S_k$ is invertible as $k\to 0$.

Let
$$
G_k^0(z)=-\frac{1}{2\pi} \ln|z| -\frac{\gamma}{2\pi}-\frac{1}{2\pi}\ln|k|,
$$
where $\gamma$ is the Euler constant, and let $S_k^0:H^{-\frac{1}{2}}(\partial \mathcal O)\to H^{\frac{1}{2}}(\partial \mathcal O)$ be the single layer operator (similar to (\ref{skk})) defined by the kernel $G_k^0(z)$:
\begin{equation}\label{sk0}
S_k^0 \sigma(z)=-\frac{1}{2\pi}\int_{\partial \mathcal O}G_k^0(z-z')\sigma (z')dl_{z'}, \quad z\in \partial \mathcal O.
\end{equation}

 Let us denote by $H^{-\frac{1}{2},\bot}(\partial \mathcal O)$ and $ H^{\frac{1}{2},\bot}(\partial \mathcal O)$ the linear subspaces in spaces $ H^{\pm\frac{1}{2}}(\partial \mathcal O)$, respectively, that consist of functions $\varphi$ such that $\int_{\partial\mathcal O}\varphi ds=0$. Every element $\psi\in H^{\pm\frac{1}{2}}(\partial \mathcal O)$ can be uniquely presented as a vector $\left(
                                                                                                                                                                       \begin{array}{c}
                                                                                                                                                                         c \\
                                                                                                                                                                         \varphi \\
                                                                                                                                                                       \end{array}
                                                                                                                                                                     \right)
 $, where $c=\int_{\partial\mathcal O}\psi ds/|\partial\mathcal O|,~\varphi=\psi-c$. These components of $\psi$ are orthogonal only in $L_2(\partial\mathcal O)$, but the norms in the original Sobolev spaces are obviously equivalent to the corresponding Hilbert-Schmidt norms, i.e., $\|\psi\|\sim(|c|^2+\|\varphi\|^2)^{1/2}$.

Using these vector representations of Sobolev spaces $ H^{\pm\frac{1}{2}}(\partial \mathcal O)$, we will write operators $S_k, S_k^0$ in the matrix form. In particular,
\begin{equation}\label{matrix}
S_k^0=\left(
                 \begin{array}{cc}
                   -\nu(\frac{\gamma}{2\pi}+\frac{1}{2\pi}\ln|k|)& b_1\\
                  b_2 & B \\
                 \end{array}
               \right)=\left(
                 \begin{array}{cc}
                   \varepsilon^{-1}& b_1\\
                  b_2 & B \\
                 \end{array}
               \right),
\end{equation}
where $B:H^{-\frac{1}{2},\bot}(\partial \mathcal O)\to H^{\frac{1}{2},\bot}(\partial \mathcal O)$ is the single layer operator (similar to (\ref{sk0})) with the kernel $-\frac{1}{2\pi}\ln|z-z'|,~\nu=|\partial\mathcal O|$, and operators $b_1,b_2,B$ are bounded and $k$-independent. Operator $B$ is a pseudo-differential operator of order $-1$. From the standard potential theory, it follows that operator $B^{-1}$ is bounded.

It was proved in \cite{{siltanen2}} that $N(kz):=G_k-G_k^0$ is an infinitely smooth function of $kz$ and $N(0)=0$. The same letter $N$ will be used also to denote the operator with the integral kernel $N(k(z-z'))$, i.e.,
\[
N:=S_k-S_k^0:H^{-\frac{1}{2}}(\partial \mathcal O)\to H^{\frac{1}{2}}(\partial \mathcal O).
\]
Then $\|N\|=O(|k|)$ as $k\to 0$. The norm is exponentially small in $\varepsilon, ~\varepsilon\to 0$, and can be estimated by $C_n\varepsilon^n$ with arbitrary $n>0$. Hence the following matrix representation is valid for  $S_k$ as $\varepsilon\to 0$:
\begin{equation*}\label{matrix1}
S_k=\left(
                 \begin{array}{cc}
                   \varepsilon^{-1}& b_1\\
                  b_2 & B \\
                 \end{array}
               \right)+\left(
                 \begin{array}{cc}
                   O(\varepsilon^n)& O(\varepsilon^n)\\
                   O(\varepsilon^n) & O(\varepsilon^n) \\
                 \end{array}
               \right).
\end{equation*}
Let $D$ be the diagonal matrix with elements $\varepsilon^{-1}, B$ on the diagonal. We multiply the equality above from the left by $DD^{-1}$. This leads to
\begin{equation}\label{matrix22}
S_k=\left(
                 \begin{array}{cc}
                   \varepsilon^{-1}& 0\\
                  0 & B \\
                 \end{array}
               \right)\left[\left(
                 \begin{array}{cc}
                   I& \varepsilon b_1\\
                  B^{-1}b_2 & I \\
                 \end{array}
               \right)+\left(
                 \begin{array}{cc}
                   O(\varepsilon^n)& O(\varepsilon^n)\\
                   O(\varepsilon^n) & O(\varepsilon^n) \\
                 \end{array}
               \right) \right], \quad \varepsilon\to 0.
\end{equation}

The second factor on the right is an operator in the space $H^{-\frac{1}{2}}(\partial \mathcal O)$.
We can use the Hilbert-Shmidt norms of all the the matrices, and they will be equivalent to the norms of the operators that are represented by these matrices. Obviously, the second factor in (\ref{matrix22}) is a small perturbation of the invertible matrix $\left(
                 \begin{array}{cc}
                   I& 0\\
                  B^{-1}b_2 & I \\
                 \end{array}
               \right)$.
Thus \begin{equation}\label{matrix122}
(S_k)^{-1}=\left[\left(
                 \begin{array}{cc}
                   I& 0\\
                  -B^{-1}b_2 & I \\
                 \end{array}
               \right)+ O(\varepsilon)
               \right]\left(
                 \begin{array}{cc}
                   \varepsilon & 0\\
                  0 & B^{-1} \\
                 \end{array}
               \right)=\left(
                 \begin{array}{cc}
                   \varepsilon & 0\\
                  0 & B^{-1} \\
                 \end{array}
               \right)+\left(
                 \begin{array}{cc}
                   O(\varepsilon^2)& O(\varepsilon)\\
                   O(\varepsilon) & O(\varepsilon) \\
                 \end{array}
               \right),
\end{equation}
where $\varepsilon\to 0$ and the remainder terms are infinitely smooth in $\varepsilon$. The invertibility of $S_k$ when $0\neq |k|\ll 1$ is proved.

From the latter fact it follows that the set $\mathcal K$ where operator $S_k$ is not invertible is a real analytic variety. Indeed, operator $S_k^0$ is an elliptic PDO of order $-1$ on the compact manifold $\partial \mathcal O$, and therefore it has zero index. Then the same is true for the operator $S_k$, since function $G_k-G_k^0$ is infinitely smooth in $(z,k)$ and analytic in $k_1,k_2$ where $k=k_1+ik_2\neq 0$ (eg \cite{siltanen2}). Thus $S_k$ is a Fredholm family of operators analytic in $k_1,k_2\neq (0,0)$.  Hence if $S_k$ is invertible at one point $k\neq 0$, then the set of  values of $ (k_1,k_2) \in \mathbb C^2\backslash\{0\}$ for which $S_k$ has a non-trivial kernel is a 1-D analytic variety (see \cite[Th.4.11]{kuchment}). The intersection of this variety with the real plane is a real analytic variety.

Let us show that $\mathcal E_D=\mathcal K$.
Let operators $\widehat{S}_k, ~\widehat{S}_k^0:H^{-\frac{1}{2}}(\partial \mathcal O) \rightarrow W^{1,p}(\mathbb R^2 \backslash \overline{\mathcal O}),~p>2,$ be the single layer operators defined by the same formula as operators $S_k, ~S_k^0$ in (\ref{skk}), (\ref{sk0}), respectively, but for all $z\in\mathbb R^2 \backslash \mathcal O$.
Consider the problem
\begin{equation}\label{ccc}
-\Delta u =0,   ~ z\in \mathbb R^2\backslash \overline{\mathcal O}; \quad u|_{\partial \mathcal O}=f \in H^{\frac{1}{2}}(\partial \mathcal O);
 \quad e^{-i\zeta \cdot z} u \in W^{1,p}(\mathbb R^2 \backslash \overline{\mathcal O}), ~ p>2.
\end{equation}
Let $k=k'\notin \mathcal K$. Then operator $S_{k'}$ is onto, and there is a function $\mu\in H^{-\frac{1}{2}}(\partial \mathcal O)$ such that $S_{k'}\mu=f$. Thus $u=\widehat{S}_{k'}\mu $ is a solution of (\ref{ccc}). If this solution is unique, then operator $F^{out}$ is well defined (by $F^{out}f=u_\nu|_{\partial \mathcal O}$) and $k'\notin \mathcal E_D$. If solution $u=\widehat{S}_{k'}\mu $ of (\ref{ccc}) is not unique, then there exists a non-trivial solution $u$ of the homogeneous problem (\ref{ccc}) when $k=k'$. Denote by $v$ the extension of $u$ by zero in $\mathcal O$. Then
\begin{equation}\label{aaa}
-\Delta v =\alpha \delta (\partial\mathcal O), ~z\in \mathbb R^2,  \quad e^{-i\zeta \cdot z} v \in W^{1,p}(\mathbb R^2), ~ p>2,
\end{equation}
where $\delta (\partial\mathcal O)$ is the delta-function on $\partial\mathcal O$ and $\alpha=u_\nu|_{\partial\mathcal O}$. From the Nachman uniqueness result \cite[Lemma 1.3]{Nachman}, it follows that $\alpha\not \equiv 0$ (otherwise $v\equiv 0$) and that $v=\widehat{S}_{k'}\alpha\delta (\partial\mathcal O)$. Thus $0=u|_{\partial\mathcal O}=S_{k'}\alpha$.
This contradicts the assumption that $k'\notin \mathcal K$. Thus $k'\notin \mathcal E_D$.

Assume now that $k=k'\in \mathcal K$. Then there exists a non-trivial $\mu$ such that $S_{k'}\mu=0. $ Function $v=\widehat{S}_{k'}\alpha\delta (\partial\mathcal O)$ is a solution of (\ref{aaa}). Function $v$ vanishes in $\mathcal O$ since $v$ is harmonic there and $v=0$ on $\partial\mathcal O$. Since the jump of the normal derivative of $v$ is proportional to $\mu$, function $v$ is not identically equal to zero. Thus $v$ is a non-trivial solution of homogeneous ($f=0$) problem  (\ref{ccc}). Thus  $k'\in \mathcal E_D$. Hence $\mathcal K=\mathcal E_D$.
To complete the proof of the first statement of Lemma \ref{lemma0408C}, it remains to recall that operator $S_k$ has zero index, and therefore it is onto when the kernel is trivial.

Let us prove the second statement of the Lemma. The following  simple formula from the potential theory is valid:
\begin{equation}\label{sf}
(F_0-F^{out})S_k=I.
\end{equation}
This formula implies that
\begin{equation}\label{ffout}
F_0-F^{out}=(S_k)^{-1}, \quad k\notin \mathcal E_D\bigcup\{0\}.
\end{equation}
Since the right-hand side is analytic in $k_1,k_2$ and $F_0$ does not depend on $k$, operator $F^{out}$ is  analytic in $k_1,k_2$ when $ k\notin \mathcal E_D\bigcup\{0\}.$ Consider the standard Dirichlet-to-Neumann map $F^{out}_b$ defined using the bounded solutions of the exterior problem for the Laplacian (the subindex ``b" here stands for ``bounded"). The difference $F^{out}-F^{out}_b$ is a smoothing operator that maps $H^{1/2}(\partial\mathcal O)$ into $H^{3/2}(\partial\mathcal O)$ (it is infinitely smoothing if $\partial\mathcal O\in C^\infty$). Thus $F^{out}$ is an elliptic pseudo-differential operator of the first order with negative symbol. The second statement is proved.

 The proof of the third statement of the lemma is based on (\ref{ffout}) and the matrix representation (\ref{matrix122}). Let us also take into account that only the lower right element of the matrix representation of the operator $F_0$ is non-zero. We will preserve the same notation $F_0$ for this element. Then
\begin{equation}\label{kvf}
F^{out}(k)=F_0-(S_k)^{-1}=\left(
                 \begin{array}{cc}
                 - \varepsilon& 0\\
                  0 &F_0 - B^{-1} \\
                 \end{array}
               \right)+\left(
                 \begin{array}{cc}
                  O(\varepsilon^2)& O(\varepsilon)\\
                 O(\varepsilon) & O(\varepsilon)\\
                 \end{array}
               \right).
\end{equation}
This formula allows us to extend $F^{out}(k)$ by continuity at $k=0$. It will completely justify the third statement of the lemma if we show that $F_0 - B^{-1}<-\delta<0$.

Let us recall that $F^{out}_b$ is the Dirichlet-to-Neumann operator that maps the Dirichlet data on $\partial\mathcal O$ into the normal derivative $u_\nu$  of the corresponding bounded solution $u$  of the exterior problem for the Laplacian. We denote by $\widetilde{S}$ the single layer operator with the kernel $-\frac{1}{2\pi} \ln|z-z'|$. Similarly to (\ref{sf}), we have that $(F_0-\widetilde{F}^{out})\widetilde{S}=I$ on functions orthogonal to constants. From here it follows that
\begin{equation}\label{kkk}
F_0-F^{out}_b=B^{-1} \quad {\rm on} \quad  H^{\frac{1}{2},\bot}(\partial \mathcal O).
\end{equation}
From the Green formula, it follows that $F^{out}_b<0$ on $H^{\frac{1}{2},\bot}(\partial \mathcal O)$. Thus $F_0- B^{-1}<0$ on $H^{\frac{1}{2},\bot}(\partial \mathcal O)$. The latter operator does not depend on $k$. It is an elliptic pseudo-differential operator of the first order (it is a restriction of $F^{out}$ to a subspace of co-dimension one), and therefore its eigenvalues tend to infinity. Thus from the negativity of $F_0 -B^{-1}$ it follows that $F_0- B^{-1}<-\delta<0$ on $H^{\frac{1}{2},\bot}(\partial \mathcal O)$.

Let us prove the last statement of the lemma. Let us recall that the kernel $G_k$ of operator $S_k$ is real-valued (see \cite[Part 3.1.1]{sthesis} and the discussion in the introduction of this paper). Thus the integral kernel of the operator $F^{out}$ is real-valued since the other two operators in (\ref{ffout}) have this property. From (\ref{ffout}) it also follows that
\begin{equation}\label{fst}
(F^{out})^\dag=-((S_k)^{-1})^\dag.
\end{equation}

In order to prove (\ref{gl}), we need to consider certain operators in Sobolev spaces with the indexes $s\in[-3/2,3/2]$, and this does not create difficulties since we assume that $\partial\mathcal O\in C^2$. We recall that $(S_k^0)^{-1}$ is a pseudo-differential operator of the first order, and from (\ref{matrix}) it follows that
\[
\|(S_k^0)^{-1}\varphi\|_{H^{-3/2}(\partial\mathcal O)}\leq C\|\varphi\|_{H^{-1/2}(\partial\mathcal O)},~~0<|k|\ll 1.
\]
It was proved in \cite{{siltanen2}} that $N(kz):=G_k-G_k^0$ is an infinitely smooth function of $kz$ and that $N(0)=0$. Hence the following estimate holds for the operator $N=S_k-S_k^0$:
\begin{equation}\label{nn}
\|N\varphi\|_{H^{3/2}(\partial\mathcal O)}\leq C|k|\|\varphi\|_{H^{-3/2}(\partial\mathcal O)}, \quad 0<|k|<1.
\end{equation}
This implies the following estimate for operator $(S_k)^{-1}=(S_k^0+N)^{-1}=(S_k^0)^{-1}(I+N(S_k^0)^{-1})^{-1}$:
 \begin{equation}\label{pr}
 \|(S_k)^{-1}\varphi\|_{H^{-3/2}(\partial\mathcal O)}\leq C\|\varphi\|_{H^{-1/2}(\partial\mathcal O)}, \quad 0<|k|\ll 1.
 \end{equation}

 Now we fix an arbitrary smooth enough $\varphi$ and denote by $\psi=\psi(k)$ the function $\psi=(S_k)^{-1}\varphi,~0<|k|\ll 1.$ Then
 \[
 (((S_k)^{-1})^\dag\varphi,\varphi)=\Im((S_k)^{-1}\varphi,\varphi)=\Im(\psi,(S_k^0+N)\psi)=\Im(\psi,N\psi),
 \]
 since operator $S_k^0$ is self-adjoint. The last equality and (\ref{nn}), (\ref{pr}) imply that
 \[
 |(((S_k)^{-1})^\dag\varphi,\varphi)|\leq |(\psi,N\psi)|\leq C|k|\|\psi\|^2_{H^{-3/2}(\partial\mathcal O)}\leq C|k|\|\varphi\|^2_{H^{-1/2}(\partial\mathcal O)},~~0<|k|\ll 1.
 \]

 Let $D:L_2(\partial\mathcal O)\to H^{-1/2}(\partial\mathcal O)$ be a bounded invertible operator. We replace $\varphi$ in the estimate above by  $D\widehat{\varphi}=\varphi$ and obtain that
 \[
  |D^*(((S_k)^{-1})^\dag D\widehat{\varphi},\widehat{\varphi})|\leq C|k|\|\widehat{\varphi}\|^2_{L_2(\partial\mathcal O)},~~0<|k|\ll 1
 \]
on a dense set in $L_2(\partial\mathcal O)$. Thus $\|D^*(((S_k)^{-1})^\dag D\|_{L_2(\partial\mathcal O)}\leq C|k|,~0<|k|\ll 1$. This and (\ref{fst}) together justify (\ref{gl}).

\qed

In order to obtain an alternative definition of the exceptional set $\mathcal E$, we reduce system (\ref{ref0401B}),(\ref{ref0401Ba}) to the boundary:
\begin{equation}\label{ref0406A}
\left \{
\begin{array}{rlll}
u &=&u^{out} + e^{i\zeta \cdot z}, & z \in \partial \mathcal O, \\
F_n u &=& F^{out} u^{out} + F_0 e^{i\zeta \cdot z}, & z \in \partial \mathcal O.
\end{array}
\right .
\end{equation}
This system immediately implies the following representation (which is equivalent to (\ref{LS})) of function $u$ at the boundary $\partial \mathcal O$:
\begin{equation}\label{LS0428}
u=(F_n-F^{out})^{-1}(F_0-F^{out})e^{i\zeta \cdot z}.
\end{equation}
The equivalence of (\ref{LS0428}) and (\ref{LS}) can be easily justified using the equality $F_n-F_0=(F_n-F^{out})+(F^{out}-F_0)$ and (\ref{sf}).

So, now we get a simple but important alternative definition of exceptional set $\mathcal E$.
\begin{theorem}\label{thkernel}
Let operator $F_n$ be well defined. Then a point $k \neq 0$ is exceptional if and only if the operator $F_n-F^{out}(k)$ has a non-trivial kernel. Moreover, the multiplicity of the exceptional point (i.e., the number of linearly independent solutions of (\ref{ref0414A})) is equal to the dimension of $Ker(F_n-F^{out}(k))$.
\end{theorem}

{\bf Remark.} If $k'\in \mathcal E_D$, i.e., the operator $F_n-F^{out}(k)$ has a singularity at $k=k'$, then the kernel is the set of functions on which both the singular and principal parts of the operator vanish. To be more rigorous, a function $\sigma$ belongs to the kernel of the operator if $\lim [F_n-F^{out}(k)]\sigma=0$ when $k\to k', ~ k\notin \mathcal E_D$.

{\bf Proof.} Let $k\in \mathcal E$ and let $\sigma=v|_{\partial\mathcal O}$, where $v$ is a non-trivial solution of (\ref{ref0414A}). From the assumption on $F_n$ it follows that $\sigma\not \equiv 0$, and equation (\ref{ref0414A}) implies that $F_n\sigma=F^{out}\sigma$. Thus $F_n-F^{out}$ has a non-trivial kernel that includes $\sigma$. Conversely, assume that $\sigma\neq 0$ belongs to the kernel of $F_n-F^{out}$ for some $k=k_0$. We define the non-trivial solution $v$ of (\ref{ref0414A}) as follows. In $\mathcal O$, it is defined as the solution of the Dirichlet problem that is equal to $\sigma$ at the boundary (recall that zero is not an eigenvalue of the interior Dirichlet problem). In $R^2\backslash \overline{\mathcal O}$, it is defined as the solution $u$ of the exterior problem (\ref{ccc1}) with $f=\sigma$ if $k_0\notin \mathcal E_D$. Otherwise, $v$ is defined as $\lim_{k\to k_0}u$. The existence of the limit follows from the Remark above.

\qed

\section{Exceptional points}

The following statement is a simple consequence of Theorem \ref{thkernel} and Lemma \ref{lemma0408C}.
\begin{theorem}
Let $n(x)$ be absorbing, i.e., $\Im n(x) \geq \delta> 0$ on $\mathcal O$. Then there are no exceptional points in a small neighborhood of the origin $k=0$.
\end{theorem}
{\bf Proof.} The Green formula implies that the quadratic form
$$
\Im( F_n u,u)=\Im \int_{\partial \mathcal O} \frac{\partial u}{\partial \nu} \overline{u}dl = \Im \int_{\mathcal O} \Delta u \overline{u}dS = - \int_{\mathcal O}\Im  n(x)|u(x)|^2 dS\leq -\delta\int_{\mathcal O}|u(x)|^2 dS
$$
is sign definite. We take into account that the standard estimates for solutions of elliptic equations are valid in the Sobolev spaces with negative indexes if the equation is homogeneous (see \cite{roitberg}). In particular, $\|u\|_{L_2(\mathcal O)}\leq C\|u\|_{H^{-1/2}(\partial\mathcal O)}$ for solutions $u$ of the equation $-\Delta u-nu=0,~x\in \mathcal O.$ Thus
\[
\Im( F_n u,u)\leq -C\delta\|u\|_{H^{-1/2}(\partial\mathcal O)}.
\]
On the other hand, Lemma \ref{lemma0408C} implies that
 \[
 \Im( F^{out} u,u)\leq C|k|\|u\|_{H^{-1/2}(\partial\mathcal O)}, \quad 0<|k|\ll 1.
 \]
Therefore, the operator $\Im(F_n-F^{out}(k))$ is sign definite for small $|k|$, and the kernel of operator $F_n-F^{out}(k),~0<k\ll 1,$ is trivial. It remains to apply Theorem \ref{thkernel}.

\qed

Let $n$ be a conductive potential vanishing outside $\mathcal O$. It means that
\[
n=-q^{-\frac{1}{2}}\Delta q^{\frac{1}{2}},
\]
where $q\in C^2(\mathbb R^2)$ is a smooth non-negative function and $q-1$ vanishes outside $\mathcal O$. Nachman proved \cite{Nachman} that there are no exceptional points for such potentials.
Perturbations $n_\lambda=n(z)+\lambda \omega(z)$ of conductive potentials were considered in \cite{siltanen2}, where $\omega$ is real-valued and supported on $\overline{\mathcal O}$. Under the assumptions that the potential is radial, i.e.,  $n=n(|z|), ~\omega=\omega(|z|)$, and
 \begin{equation}\label{1405A}
 \mu= \int_{\mathcal O} \omega q dS > 0,
 \end{equation}
the authors of \cite{siltanen2} proved that the exceptional set is empty for small negative $\lambda$, and there exists an exceptional set for positive small $\lambda$. (Formally, the sign of $\lambda$ in the latter statement is opposite to the one used in \cite{siltanen2} since here we use the wave equation with the different sign before the potential). It was shown that the exceptional set is a circle of radius $e^{-\frac{1}{\mu \lambda}(1+o(1))}, ~ \lambda \rightarrow +0$.

Our approach allows us to extend this result to the case of non-radial potentials. The exceptional set in this case is not a circle anymore, but it approaches a circle as $\lambda \rightarrow +0$. Consider the variables $\varepsilon=[-\nu(\frac{\gamma}{2\pi}+\frac{1}{2\pi}\ln|k|)]^{-1}, ~ \varphi=\arg k, ~\varphi \in [0,2\pi)$.
\begin{theorem}\label{0408E}
Let $n_\lambda=n(z)+\lambda \omega(z)$, where $n$ is a conductive (real-valued) potential, $\omega$ is real-valued, $n(z)=\omega(z)=0$ when $z\notin \overline{\mathcal O}$, and (\ref{1405A}) holds.

If $\lambda<0$ is small enough, then the exceptional set $\mathcal E$ is empty.
Moreover the following estimate holds for the scattering transform (\ref{dst}): $|t(k)|<C(\lambda)/|\ln|k||, ~ k \rightarrow 0$.

If $\lambda>0$ is small enough, then the exceptional points exist only in a neighbourhood of the origin and the exceptional set is given by the equation $\varepsilon= \mu \lambda(1+o(1)), ~ \lambda \rightarrow +0$, where the remainder depends smoothly on $\lambda$ and $\varphi$.
\end{theorem}
{\bf Proof.}
We have (eg \cite[(3.18)]{RN2}) that
\begin{equation}\label{0410C}
|G_k(z)e^{-i\zeta \cdot z}|\leq \frac{c}{\sqrt{|k|}\sqrt{|z|}}, ~ c>0, ~ E=0.
\end{equation}
This implies the unique solvability of the Lippman-Schwinger equation
$$
u-e^{i\zeta \cdot z}= -G_k * (n_\lambda u)
$$
when $|n_\lambda|<C$ and $|k|$ is large enough. Problem (\ref{ref0414A}) has only trivial solution for these $k$ and $\lambda$. Hence there exists $K_0>0$ such that the region $|k|>K_0$ is free of points $k\in \mathcal E$ when $|n_\lambda|<C $ (see more details in \cite[proof of the corollary 3.5]{siltanen2}).

Now we are going to show that the exceptional points for potential $n_\lambda$ may occur only in a small neighborhood of the origin $k=0$. Indeed, operator $F_n$ is a pseudo-differential operator of the first order with a positive principal symbol. Due to Lemma \ref{lemma0408C}, operator $F^{out}(k),k\neq 0,$ is a pseudo-differential operator of the first order with a negative principal symbol. Hence, $F_{n}-F^{out}(k)$ is an elliptic operator of the first order, and therefore, its eigenvalues tend to infinity. We take additionally into account that the kernel of the operator $F_{n}-F^{out}(k)$ is trivial for all $k\neq 0$ due to Theorem \ref{thkernel}. This implies that the operator $(F_{n}-F^{out}(k))^{-1}$ is bounded for each fixed $k\neq 0$. From the analyticity in $k_1,k_2$ it follows that the upper bound for the norm $\|(F_{n}-F^{out}(k))^{-1}\|$  can be chosen uniformly in $k$ on each region of the form $K_0\geq |k|\geq \delta>0$. Then the same is true if $n$ is replaced by $n_\lambda$ with small enough $|\lambda|$. Hence Theorem \ref{thkernel} implies that the exceptional points for the problem with the perturbed potential $n_\lambda$ and sufficiently small $|\lambda|$ can appear only in a small neighborhood of $k=0$.

Now let us study the structure of the set $\mathcal E$ in a neighborhood of the origin $k=0$. Since the substitution $u=\sqrt{q}v$ reduces equation (\ref{ref0401B}) with a conductive potential  to the equation $\nabla (q \nabla) v=0$, the D-t-N maps for these equations coincide. Hence, the kernel and co-kernel of operator $F_n$ are one dimensional spaces of constants. The norm of the restriction of $F_n$ on the space $L^{2,\bot}$ of functions orthogonal to constants is greater than some positive constant.


Consider the operator $A(\lambda,k):=F_{n_\lambda}-F^{out}(k)$. From Lemma \ref{lemma0408C} and the properties of $F_n$ established above, it follows that $A(0,0)$ has zero eigenvalue with constant eigenfunction, and all the other eigenvalues are greater than some positive constant $\delta>0$. Operator $F_{n_\lambda}$ is analytic in $\lambda$, and operator $F^{out}(k)$ is an infinitely smooth function of $\varepsilon=[-\nu(\frac{\gamma}{2\pi}+\frac{1}{2\pi}\ln|k|)]^{-1}$ at $\varepsilon=0$ with all the derivatives at $\varepsilon=0$ independent of the polar angle of $k$. The properties of  $F^{out}(k)$ are proved in Lemma \ref{lemma0408C}. In fact, the independence of the derivatives of $\varphi$ is not stated there, but could be easily verified in the process of the proof. Hence operator $A(\lambda,k)$ with small enough $|\lambda|+|k|$ has an eigenvalue $\xi$ of the form
$$
\xi(\lambda,\varepsilon,\varphi)=a \lambda + b\varepsilon + O(\lambda^2+\varepsilon^2)
$$
with  a smooth in $k,\lambda$ eigenfunction $e(k,\lambda)$ and all the other eigenvalues being separated from zero. The latter statement for general analytic families of operators with an isolated eigenvalue can be found in \cite[XII.8]{reed}. One can easily see that the proof there does not require the analyticity and remans valid for smooth operator functions.

Let us find constants $a$ and $b$.  Let $e=e(0,0)$. Recall that $e$ is a constant. We normalize $e(k,\lambda)$ in such a way that $e\equiv 1$. Operator $A(0,0):=F_{n}-F^{out}(0)$ is self-adjoint, and therefore
\begin{equation}\label{2511A}
a=(A(\lambda,k)e(\lambda,k),e(\lambda,k))'_\lambda(0,0)=\left ( \frac{\partial}{\partial\lambda}A (0,0)e,e \right )=\left (\frac{\partial}{\partial\lambda}F_{n_\lambda}(0,0)e,e \right ),
 \end{equation}
where $e=e(0,0)$. We used here that,
$$
(A(0,0)e'(0,0),e) = (A(0,0)e,e'(0,0)) = 0,
$$
since $A(0,0)$ is self-adjoint and $A(0,0)e=0$.

Let us evaluate the right-hand side in (\ref{2511A}).
Consider solutions $f_\lambda \in H^{1/2}(\mathcal O)$ of the equation  $\Delta f_\lambda + n_\lambda f_\lambda =0$ in $\mathcal O$ subject to the boundary condition $f_\lambda  = e$ at $\partial \mathcal O$. Note that its derivative satisfies $\Delta f'_\lambda + n_\lambda f'_\lambda= - n'_\lambda f_\lambda$ in $\mathcal O,~f_\lambda' =0$ at $\partial \mathcal O$. From the Green formula it follows that
\[
\int_{\partial O} \frac{\partial f_\lambda '}{\partial \nu} \overline{f} dl  =\int_{\mathcal O} n'_\lambda |f_\lambda|^2 dS .
\]
We put here $\lambda=0$ and take into account that $f_\lambda =q^{1/2}$ when $\lambda=0$. This leads to
$$
\left (\frac{\partial}{\partial\lambda}F_{n_\lambda}(0,0)e,e \right ) = -\mu,
$$
where $\mu$ is given by (\ref{1405A}). Hence $a=-\mu$.

Similarly, from (\ref{kvf}) it follows that
\[
b=(A e,e)'_\varepsilon(0,0)=(A_\varepsilon(0,0)e,e)=-\left (\frac{\partial}{\partial\varepsilon}F^{out}(k)e,e \right )|_{\varepsilon=0}=1.
\]
Thus
\begin{equation}\label{aaaa}
\xi(\lambda,\varepsilon,\varphi)=-\mu \lambda + \varepsilon + O(\lambda^2+\varepsilon^2), ~ |\lambda|+|k|\ll 1,~~\mu>0.
\end{equation}

Since the set $\mathcal E$ is located in a small neighborhood of the origin $k=0$, from Theorem \ref{thkernel} it follows that $\mathcal E$ is defined by the relations $\xi(\lambda,\varepsilon,\varphi)=0, ~0<\varepsilon\ll 1$. Since $\varepsilon>0$, all the statements of the theorem that do not concern $t(k)$ follow immediately from (\ref{aaaa}).

Now let us estimate $t(k)$ in a neighbourhood of $k=0$. From (\ref{dst}) and (\ref{LS0428}) it follows that $|t(k)|$ can by estimated by $C\|(F_{n_\lambda}-F^{out})^{-1}(F_0-F^{out})e^{i\zeta \cdot z}\|$. From Theorem \ref{thkernel} it follows that for $\lambda < 0$ and small $|\lambda |$,  operator $(F_{n_\lambda}-F^{out})^{-1}$ is bounded in the small neighbourhood of $k$. Thus
$$
|t(k)| \leq C(\lambda) \|(F_0-F^{out}(k))e^{i\zeta \cdot z}\|,~ |k|\ll 1.
$$
It remains to use representation (\ref{matrix122}) for operator $F_0-F^{out}(k)$ (see also (\ref{ffout})) and write $e^{i\zeta \cdot z}$ in the form $1+O(k)$.

 \qed.

\section{Condition for existence of exceptional points}
Now we present a method that allows one, in some cases, to justify existence of exceptional points on a path $\gamma\subset \mathbb C$ that is analytic in $k_1,k_2$ by making certain measurements at the end points of $\gamma$. In this section, we assume that the potential $n$ is real-valued.

Consider the operator function $P(k):=I+S_k(F_n-F_0)$ in $L_2(\partial\mathcal O)$ or $H^{-1/2}(\partial\mathcal O)$. The integral kernel of $P(k)$ is real-valued, since function $G_k$ is real-valued (see \cite[Part 3.1.1]{sthesis}). Hence if $\mu$ is an eigenvalue of  $P(k)$, then the complex conjugate number $\overline{ \mu}$ is also an eigenvalue of the same multiplicity. We already mentioned earlier that operator $S_k(F_n-F_0)$ is compact. Thus for each $k$, the eigenvalues $\mu_i=\mu_i(k)$ converge to $\mu=1$ as $i\to\infty$ and therefore, $P(k)$  has at most a finite number of negative eigenvalues.
We can  introduce a function that counts the number of negative eigenvalues $\mu_i(k)$ of operator $P(k)$:
$$
n^-(k)= \sum_{i ~:~ \mu_i(k)<0} m_i,
$$
where $m_i$ is the algebraical multiplicity of $\mu_i$. In the case of positive energy, operator function $P_E(\lambda)$ and the counting function $n^-(\lambda)$ are introduced absolutely similarly to the corresponding objects in the case of $E=0$.

\begin{theorem}\label{t22}
 Let energy be zero, and
let $k,\widehat{k} \in \mathbb C \backslash \{0\}$ be arbitrary points such that
$n^-(k)\neq n^-(\widehat{k}) ~(mod ~2)$. Then every analytic path $\gamma$ connecting points $k$ and $\widehat{k}$  and not passing through $k=0$ contains at least one exceptional point.

The same statement (with $k,\widehat{k}$ replaced by $\lambda, \widehat{\lambda}$) is valid when energy is positive under additional condition that the path $\gamma$ does not contain points of the  unit circle $|\lambda|=1$.
\end{theorem}

{\bf Proof.}
  Let $E=0$, and let $k=k(s),~0\leq s\leq 1,$ be an analytic parametrization of $\gamma$. Then operator $P(k(s))$ is analytic in $s$. Since operator $S_k(F_n-F_0)$ is compact, the spectrum of $P(k(s))$ consists of eigenvalues $\mu_i(k(s))$ of finite multiplicities. Thus (see \cite[consequence of Th.XII.2]{reed}) the eigenvalues $\mu_i(k(s))$ are analytic in $s$ except for a possible finite amount of branching points. Moreover, for each eigenvalue $\mu_i(k(s_0)),~0\leq s_0\leq 1,$ there exists a complex neighborhood $V$ of the eigenvalue and $\varepsilon>0$ so small that the number of all eigenvalues $\mu_i(k(s))$ in $V$ with a fixed $s,~|s-s_0|<\varepsilon,$ counted with their algebraical multiplicity does not depend on $s$. Therefore, function $n^-(k(s))$ can change its value (when $s$ changes from $0$ to $1$) only if an eigenvalue $\mu(k(s))$ of $P$ passes through the point $\mu=0$ or if some eigenvalues leave/come to the real axis. The second option occurs only with pairs of complex-adjoint eigenvalues. Therefore, $n^-(k(1))-n^-(k(0))$ can be odd only if at least one of the eigenvalues  $\mu(k(s))$ passes through $\mu=0$. The corresponding point $k(s)\in\gamma$ is exceptional. The statement of the theorem in the case $E=0$ is proved.

The proof in the case $E>0$ remains the same. The additional condition that $\gamma$ does not intersect the unit circle is needed only because the parameter $\lambda$ in the Faddeev scattering problem with positive energy can't belong to the unit circle.

 \qed

{\bf Acknowledgments.} The authors are thankful to Eemeli Bl\aa sten, Uwe Kahler, Michael Music, Roman Novikov, and Samuli Siltanen  for useful discussions concerning the Faddeev scattering problem. Authors are thankful to the anonymous referee who found an essential error in the previous version of the article.

\end{document}